# US MUON WORKSHOP 2021

*A road map for a future Muon Facility*

Despina Louca
Gregory J. MacDougall
Travis J. Williams

**Date: Sep. 22, 2021**

**Co-sponsors:**

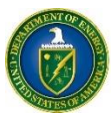 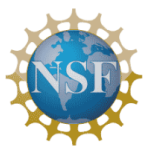

**Report from:**
**US Muon Workshop 2021: A road map for a future Muon Facility**

**February 1-2, 2021**

Despina Louca
*University of Virginia*

Gregory J. MacDougall
*University of Illinois Urbana-Champaign*

Travis J. Williams
*Oak Ridge National Laboratory*

Draft Date:
September 22, 2021

This manuscript has been authored in part by UT-Battelle, LLC, under contract DE-AC05-00OR22725 with the US Department of Energy (DOE). The publisher acknowledges the US government license to provide public access under the DOE Public Access Plan (http://energy.gov/downloads/doe-public-access-plan).

# CONTENTS

# ACRONYMS

| | |
|---|---|
| μSR | Muon spin rotation/relaxation/resonance |
| AFM | Antiferromagnet |
| ALC | Avoided level crossing |
| DFT | Density functional theory |
| FTS | First target station |
| ISIS | ISIS Neutron and Muon Facility |
| J-PARC | Japan Proton Accelerator Research Complex |
| LE-μSR | Low-energy μSR |
| MBE | Molecular beam epitaxy |
| MHz | Megahertz ($10^6$ Hz) |
| NMR | Nuclear magnetic resonance |
| ORNL | Oak Ridge National Laboratory |
| PPU | Proton power upgrade |
| PSI | Paul Scherrer Institute |
| QENS | Quasi-elastic neutron scattering |
| REXS | Resonant x-ray scattering |
| RF | Radio frequency |
| SANS | Small angle neutron scattering |
| SC | Superconducting |
| SEE | Single event effects |
| SEEMS | Single Event Effects & Muon Spectroscopy |
| SNS | Spallation Neutron Source |
| STS | Second target station |
| THz | Terahertz ($10^9$ Hz) |
| TRIUMF | TRIUMF Meson Facility |
| US | United States |
| USD | United States dollars |
| β-NMR | β-detected Nuclear Magnetic Resonance |

# EXECUTIVE SUMMARY

The workshop titled "US Muon Workshop 2021: A road map for a future Muon Facility" was held virtually on February 1-2, 2021. The workshop aimed to bring together world experts in muon spectroscopy (µSR) and other techniques along with interested stakeholders to evaluate the scientific need to construct a new µSR facility in the United States (US). The more than 200 participants highlighted several key scientific areas for µSR research, including quantum materials, hydrogen chemistry, and battery materials, and how each area could benefit from a new, high flux pulsed muon source. Experts also discussed aspects of the µSR technique, such as low-energy µSR, novel software developments, and beam and detector technologies that could enable revolutionary advances in µSR at a next-generation facility. The workshop concluded with discussion of a concept being developed for a new µSR facility at the Spallation Neutron Source (SNS) of Oak Ridge National Laboratory (ORNL). That novel design concept was first envisioned by many of the same µSR experts at a workshop held previously at ORNL in 2016. The participants expressed that the current design had the potential to be a world-leading µSR facility, and strongly encouraged the principal investigators to continue their work in order to refine the concept and determine instrument parameters that would enable new scientific opportunities.

Muon Spin Rotation/Relaxation/Resonance (µSR) is a technique that involves the use of spin-polarized muons that are implanted in a material to provide extremely sensitive measurements of the local magnetic field distribution within samples of scientific interest. The µSR technique has led to important results in condensed matter physics, chemistry and semiconductor physics, among other fields. This technique is highly complementary to neutron scattering and since the two techniques share a common user base, 3 of the 4 existing µSR facilities in the world are co-located with neutron sources. The exception is in North America, where the sole muon source is located at a meson accelerator laboratory in Vancouver, Canada. The United States has not had a µSR facility since the closure of LAMPF at Los Alamos National Laboratory, and never one that was globally competitive. Accordingly, there have been several efforts in recent years to address this shortcoming, most recently at ORNL beginning in 2016, and culminating with this workshop.

Several recurring themes were identified during the workshop: the advantage of higher muon fluxes to enable new science, increasing demand for low-energy muon beams, the need for more software tools for muon site determination and analysis, and the role of multi-probe studies combining µSR with neutrons and other spectroscopic techniques. The primary method for enabling new science with µSR is higher flux muon beams. It is important for the detection of weak magnetic field phenomena, delivers greater sensitivity to molecular levels, and even facilitates broader applications such as using muon beams for fundamental physics experiments. But by far the largest benefit of a high muon flux would be the expansion of low-energy µSR capabilities. Low-energy µSR beams enable more depth-resolved experiments, creating opportunities for measuring topological materials, novel states in interfaces, layered heterostructures and other new types of experiments. In particular, there is an opportunity to focus the low-energy muons into a sub-millimeter beam to create a muon microscope for adding spatial resolution. The consensus, based on recent history and the state of the community, was that any opportunity to expand the capabilities of low-energy µSR would be hugely beneficial to the scientific community.

The workshop often noted the complementary nature of µSR to other techniques, especially neutron scattering. Researchers always benefit from having access to other types of measurements of the materials. The co-location of µSR and neutron scattering facilities has proven this fact, and a next-generation muon source in the US would miss scientific opportunities by not being closely associated with existing national expertise in the areas of neutron scattering, computing, advanced materials characterization and other spectroscopic techniques. Co-locating these facilities at a world-class institution like ORNL and fostering collaborations between closely-aligned communities would be a great benefit to scientific discovery, materials engineering and technological development. These key features are central to the developing plans for a US-based µSR facility located at the Spallation Neutron Source at ORNL. The concept in development has the opportunity to gain orders of magnitude more flux, and it was widely agreed among workshop participants that it would be an excellent opportunity to establish new low-energy µSR research instruments in the US that can expand the current capabilities of the technique, while pushing the bounds of scientific discovery. The µSR community expressed strong interest in being further engaged in realizing a new, US-based muon source.

# INTRODUCTION

Muon Spin Rotation/Relaxation/Resonance (µSR) is a technique that involves using spin-polarized muons implanted in a material to provide extremely sensitive measurements of the static and dynamic properties of the local magnetic field distribution within samples of scientific interest. Proton beams are directed into a target (typically Carbon or Beryllium) to produce pions. The pions decay with a mean lifetime of 26 ns via the weak interaction into a muon (or antimuon) and an anti-muon neutrino (or muon neutrino). Muons produced from pions at rest (residing on or near the surface of the target) are known as surface muons and are the most frequently used muons for µSR experiments. The muon carries most of the momentum from this decay but will stop in thin samples, with a stopping range given by ~120 mg/cm$^2$ over the sample density (~1/3 mm in carbon). Conversely, muons produced from pions that have been ejected from the target are known as decay muons. These decay muons are useful because of their higher momentum, resulting in deeper penetration into samples and allowing for measurement of materials in an enclosure or high-pressure environment. Finally, beams of low energy muons can be produced by taking beams of surface muons and reducing their momentum. This is currently done on a single active user beamline by passing the beams through thin sheets of noble gases at low temperatures, but there is a developing effort to slow muons using laser pumping. Due to the sharp drop in the muon flux during the slowing process (4-5 orders of magnitude), a high initial flux is needed to make beams of low-energy muons useful for materials research. Beams of low-energy muons have a much shorter stopping distance, allowing measurements of thin films, nanostructures and surface properties. Because the pion decay is governed by the weak interaction, which violates parity, both the neutrinos and muons produced are exclusively left-handed; that is, their spin is antiparallel to their linear momentum in the rest frame of the pion. Thus, the surface muons will be 100% spin polarized, while decay muons have marginally lower (>80%) spin polarization since finite momentum ranges of the pion and muon must be allowed, in order to give sufficient flux. Depending on the charge of the pion, either positive or negative muons can be produced and used to perform µSR experiments. However, since negative muons have more complex interactions with the samples being measured, nearly all µSR experiments in current facilities are performed with positively-charged antimuons. For the remainder of the report, "muons" could refer to either negatively- or positively-charged muons, but for practical purposes, we strongly emphasize the use of positively-charged muons.

Experiments exploiting the unique properties of µSR have led to important results in condensed matter physics, chemistry, and semiconductor physics, among other fields. In the study of magnetism especially, this technique is highly complementary to neutron scattering: where neutrons provide bulk measurements in reciprocal space with fluctuations on the timescale of nanoseconds (THz), µSR is a local, real space probe, sensitive to microsecond-scale fluctuations (MHz). The two techniques share a common user base, and for this reason 3 of the 4 existing µSR facilities in the world are co-located with neutron sources (PSI, ISIS and J-PARC). The glaring exception is in North America, where the sole muon source is located at the meson accelerator laboratory TRIUMF in Vancouver, Canada. The United States has never developed a competitive facility. To address this national shortcoming, there have been several efforts in recent years to assess the feasibility of building a US facility for µSR. This includes a conversation during the construction of the first target station (FTS) of the Spallation Neutron Source (SNS) in 2000 and, more recently, meetings held in the context of ProjectX at FermiLab (2013) and the Transformative Hadron Beamlines initiative at Brookhaven National Laboratory (2014). At Oak Ridge National Laboratory (ORNL), recent successes at the SNS, along with the ongoing Proton Power Upgrade (PPU) and Second Target Station (STS) expansion projects, have motivated a larger conversation about future science possibilities at the lab. These conditions and the clear national interest created strong motivation to explore the feasibility of a µSR source co-located with the SNS, leading to a gathering of experts in the fields of muon, neutron and accelerator science at ORNL on September 1-2, 2016. The workshop, entitled "Future Muon Source Possibilities at the SNS", lead to the inception of a novel source design concept, which has been further refined in scientific and technical scope in subsequent years.

This led Despina Louca (University of Virginia), Greg MacDougall (University of Illinois Urbana-Campaign) and Travis Williams (Oak Ridge National Laboratory) to convene this workshop, aimed at gathering input from the µSR and wider scientific communities on the needs for a next-generation µSR facility in the United States. The workshop featured seven talks given by µSR experts in a range of scientific and technical areas, who were specifically asked to consider the current limitations and future directions of their field. This led to four breakout sessions, chaired by members of the community, to discuss the role that a next-generation µSR facility could play in future scientific discovery. It concluded with an update on the progress of the µSR facility concept at ORNL and a discussion of the future work planned on this project. This report summarizes the presentations from the workshop in Section 1. Section 2 summarizes the conclusions from the breakout sessions, including the conceptual overview for a muon source and a comparison of the relative figures of merit to other sources around the world. The report concludes with a summary of main outcomes from the workshop and potential paths forward.

# 1. PRESENTATION SUMMARIES

## 1.1 YASUTOMO UEMURA, COLUMBIA UNIVERSITY – 'MUON SPIN RELAXATION (μSR) AND NEUTRON SCATTERING STUDIES IN QUANTUM MATERIALS.'

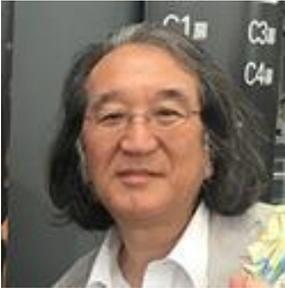

Prof. Uemura, a Condensed Matter Experimentalist from Columbia University, opened the workshop with the first talk on quantum materials with muons and neutrons. Prof. Uemura is the inaugural recipient of the Toshimitsu Yamazaki prize for outstanding, sustained work in μSR with long term impact on science. His foundational work in the areas of spin glasses, quantum magnetism and unconventional superconductivity have given him a unique perspective on the power of muon techniques and their impact on condensed matter research. His talk addressed a wide variety of condensed matter systems with interesting magnetic properties, drawing attention between the complementarity of neutron scattering, NMR and muon spectroscopy. These are listed below:

In studies of quantum materials, Muon Spin Relaxation (μSR) measurements can provide important information complementary to neutron scattering studies in the following aspects: (1) Different time windows: μSR has a sensitivity to the spin fluctuation rate ranging between $10^6$ to $10^{11}$ s$^{-1}$, covering regions with slower fluctuations (lower energy transfers) than neutrons. Examples included studies of dilute alloy spin glasses [1], $NiGa_2S_4$ [2], and MnSi "partial order" [3]. (2) μSR can detect a very small static magnetic moment, of the size of nuclear dipolar moments even in highly disordered / random spin configurations. Examples included time reversal symmetry breaking in $Sr_2RuO_4$ [4], and $UPt_3$ [5], as well as details of spin glasses [1]. (3) Neutron scattering Bragg peak intensity is proportional to the ordered moment S squared multiplied by the ordered volume fraction $V_M$. μSR can provide independent information on the local ordered moment size and $V_M$. This feature helps with the detection of phase separation and first order magnetic transitions. Examples included studies of Mott transition systems $V_2O_3$, $RENiO_3$ [6], MnSi tuned by hydrostatic pressure [3], phase boundaries between parent antiferromagnetic (AFM) and superconducting (SC) states in unconventional superconductors [7]. (4) Absence of static magnetic order can be confirmed with much better accuracy by μSR as compared to neutron scattering. μSR's sensitivity to slow spin fluctuations is particularly helpful with this. Examples included quantum spin liquids [8,9], low dimensional spin systems [10], and frustrated magnets. (5) The magnetic field penetration depth of superconductors can be determined by μSR. The energy scales inferred from the superfluid density from μSR can be combined with the energy scales of the magnetic resonance mode from neutron scattering. Examples included high-Tc cuprate, FeAs, and heavy-fermion superconductors [11]. μSR can be applied to thin films with the thickness of 200 Angstroms or more, as well as provide essential information even with polycrystalline or powder samples. The amount of specimens required is about 100 mg, which is significantly less than in neutron scattering. Examples included organic conductors [12,13], and $C_{60}$ systems [14].

Prof. Uemura further emphasized that when new magnetic materials are synthesized, it would be most sensible to perform μSR first, followed by more detailed studies of spin structures and spin excitations by neutron scattering. In his presentation, he emphasized the significant merits of performing μSR and neutron scattering on the same materials and compare and combine their results. Having both capabilities at one facility would lead to quite productive studies of novel magnetic/superconducting quantum systems.

## 1.2 JUN SUGIYAMA, CROSS NEUTRON SCIENCE AND TECHNOLOGY CENTER – 'BATTERY MATERIALS WITH MUONS'

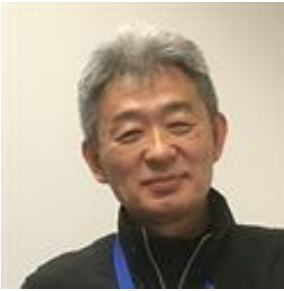

Dr. Sugiyama has a long history of using muon beams for industry-oriented research relating to battery materials. As a former researcher with Toyota Central R&D Laboratories, his expertise lies in measuring lithium and sodium ion diffusion rates in batteries. He brings tremendous insight into the industrial motivations for a next-generation muon source. His talk addressed a wide variety of battery related systems with interesting properties, drawing attention between the complementarity of neutron scattering, NMR and muon spectroscopy.

Dr. Sugiyama discussed how muon spin rotation and relaxation (μSR) is a very powerful and sensitive tool to study local magnetic environments in condensed matters, even in a zero magnetic field due to complete muon spin polarization [15]. This feature is naturally attractive for magnetic materials research, but also is essential for energy materials research, such as battery materials and hydrogen storage materials research, through the observation of a nuclear magnetic field. By using this feature, $\mu^+$SR distinguishes a nuclear magnetic field from an electron magnetic field in a paramagnetic state of a magnetic material. In fact, jump diffusion of $Li^+$ ion in $Li_xCoO_2$ was successfully measured with $\mu^+$SR [16], despite the presence of magnetic Co ions in the lattice. Note that Li-NMR is unable to detect Li diffusion in materials containing magnetic ions. Since then, many battery materials have been investigated with $\mu^+$SR in order to determine the intrinsic jump diffusion coefficient ($D^J$) of $Li^+$, $Na^+$, and $K^+$ ions [17,18]. Note that a correct $D^J$ has never been obtained with electrochemical measurements due to the absence of information on reactive surface area.

Since the mass of muon is about 1/9 of the mass of proton, one has a naive notion that muon is more mobile than $Li^+$ and/or $Na^+$ in the lattice. In order to test this hypothesis, the dynamics of a nuclear magnetic field in $LiMnPO_4$ was studied with both positive and negative muons, i.e., with $\mu^\pm$SR. Muon diffusion can only occur in $\mu^+$SR, because the implanted $\mu^-$ forms a stable muonic atom at the lattice site, and therefore any dynamics of a nuclear magnetic field measured with $\mu^-$SR must be due to Li diffusion. As a result, it was confirmed that muons are capable to probe the Li diffusion in $LiMnPO_4$ [19].

One possible future direction of μSR work on battery materials is expected to involve in-situ measurements [20], for example during an electrochemical reaction, using a high flux muon beam provided in the advanced proton accelerator like the Spallation Neutron Source (SNS) of Oak Ridge National Laboratory. In-situ measurements for μSR are possible, but rely on long counting times, the scope of which would be greatly enhanced with high muon flux beams. Recently developed multi-detector counting systems have increased the counting rate of the $\mu^+$SR spectrum that is measured per one muon pulse, i.e., 25 Hz in J-PARC [21]. Such technological advancements will enable us to measure chemical diffusion coefficient ($D^C$) of ions in battery materials with a combination of *in-situ* electrochemical measurements in the near future.

### 1.3 JAMES LORD, ISIS, RUTHERFORD APPLETON LABORATORY – 'MUON SPIN RELAXATION IN SEMICONDUCTORS AND OXIDES'

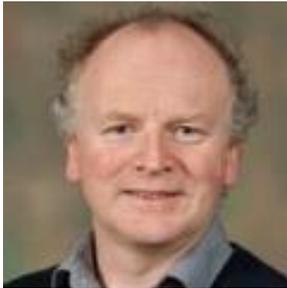

Dr. Lord is an instrument scientist on the HiFi muon Instrument at the ISIS Neutron and Muon Source. His research focuses on condensed matter systems, specifically semiconductors and oxides. Dr. Lord is also an expert on muon instrumentation and research using high magnetic field measurements with muon spectroscopy.

Dr. Lord discussed the impact of µSR for semiconductor research. Hydrogen is a ubiquitous impurity in most semiconductors because of processing, and is usually electrically active. The low concentration makes direct measurements difficult. Many technologically important oxide materials and insulators also contain hydrogen as an impurity. The positive muon ($\mu^+$) can be considered as a light isotope of hydrogen, known as Muonium (Mu), and is used as a model for hydrogen to determine its site and electrical behavior. Spectroscopy using Mu is often based on measuring the muon signal in an applied field, where an unpaired electron is either in a spin-up or spin-down state relative to the field, and adds or subtracts a local field to the muon. At low temperatures, the signals for different muonium charge states can be seen, and the fraction of each state depends on the capture rate of free electrons or holes. By measuring the relative population of the various charge states and their temperature dependence, energy barriers within the materials can be measured. Notably, hydrogen can also act as a dopant in many materials to increase conductivity. The same principle can be measured with Mu in semiconductors and oxides.

Finally, he also commented on the future directions and the development of more advanced instrumentation. The main limitation for measurements of semiconductors and oxides is muon flux and the ability to focus muon beams to small spot sizes. A next-generation source with high flux could overcome these limitations and open new avenues of research. Additionally, optical pump-probe measurements are being utilized more, where the laser pulse is timed to match the muon pulse and the penetration depth of both probes must also be matched. Signal amplitudes are limited by laser intensity, and pulse- and energy-matching costs flux, meaning that a higher flux source would be able to push the range of measurements. Another type of measurement involves mis-matching the pulse timings in order to look at carrier or excitation lifetimes, a much slower measurement with the current state of instrumentation, but one that could be more routine in a future µSR source.

### 1.4 PETER BAKER, ISIS, RUTHERFORD APPLETON LABORATORY – 'IONIC DIFFUSION.'

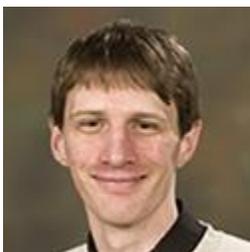

Dr. Baker is an instrument scientist at the ISIS Neutron and Muon source, which hosts pulsed muon beams for performing a wide range of scientific experiments. Dr. Baker's talk discussed his research, which combines probing ionic diffusion with magnetism and superconductivity and highlights the versatility of muon spectroscopy as a technique.

Ion jumps in materials occur on the length scale of nanometers with a frequency of MHz, which is ideally matched to be probed with µSR. This has resulted in significant research into lithium diffusion in battery materials [22], the motion of muons in materials [23] and more recently to less commonly studied ions like $Mg^{2+}$ and $I^-$ [24,25]. This can be complementary to quasi-elastic neutron scattering (QENS) that, while it is far more difficult for studies of Li ionic diffusion, can provide the momentum-dependence complement to the real-space probe of µSR.

A future muon source that is more intense offers the prospect of extending the capabilities of the technique. A higher muon flux means not only data collection at higher rates, but also increasing the range of measureable ionic diffusion rates and opens the possibility of studying more difficult nuclei and connecting with more advanced data modelling. Battery materials present tremendous opportunities for in-operando μSR measurements, which could be a new capability of a next-generation muon source. Moreover, ionic diffusion has been a topical area of research for perovskite solar cells, where there is evidence for performance hysteresis related to ion movement.

He closed by noting that the muon flux is the primary determinant of the limits of μSR measurements in chemical systems. More flux enables more dynamic range, opens more systems for study, allows for more complex computation and modeling, and as he noted, "collecting your data faster makes everything better".

## 1.5   THOMAS PROKSCHA, PAUL SCHERRER INSTITUTE – 'LOW-ENERGY MUONS'

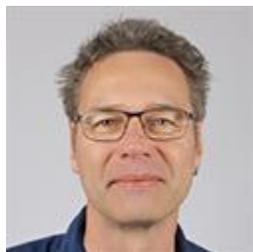

Dr. Prokscha is the Head of the Low-Energy Muon group at PSI, the world's best source for low energy muon beams. He brings an unparalleled vision for the development and future need for studies using low-energy muon beams, including to thin films, devices and surface studies.

In his talk, he began by outlining the properties of low-energy muon beams and how low-energy μSR (LE-μSR) is achieved at PSI. LE-μSR can stop muons in the range of 5-100 nm of the material surface. This can in theory be done in several different ways, and researchers are always looking for more efficient ways of producing LE-μSR beams. One option is to use degraders, which is a relatively easy approach to implement, but it broadens the energy width of the resultant beam. This translates to a loss of precision with the penetration depth of the muon. Furthermore, it contaminates the beam with muonium [26,27,28]. Another approach is to use laser ionization of muonium and is the subject of ongoing testing at J-PARC [29]. This has been shown to be reasonably efficient but reduces the muon polarization by ~50% [30]. This is potentially the best method for low-flux muon beams that require more efficient moderation, but at a new, high-flux source would not be optimal.

The method that is currently being used at PSI is to use a solid noble gas that has been solidified on a thin foil. In their case, PSI uses solid Ar on Ag foil, and achieves good moderation. This is not without a dramatic reduction in flux, obtaining approximately 1 low-energy muon per $5 \times 10^4$ surface muons [31,32,33]. The main advantage of this method that the moderation is extremely rapid (~10ps), and so there is no effect on the muon spin polarization or on the muon pulse width for pulsed beams. Other foils and gases have been proposed, and a next generation muon source would be able to design a LE-μSR beamline from scratch, aiming for the most efficient production of low-energy muons.

The technique of LE-μSR is growing rapidly in demand, having already seen applications to various surface and interface phenomena including topological materials. Interestingly, it was noted that LE-μSR is one of the best ways to attract new μSR users, since it is growing the applicability of μSR to measuring thin films and heterostructures. This parallels extremely well with neutron scattering, where there is an increasing use of neutron reflectometry for depth-resolved measurements of surface states and proximity effects. Furthermore, it was highlighted how LE- μSR has been used to probe defect effects with nanometer resolution. As a local probe, this is something to which μSR is extremely sensitive and which cannot be seen with neutrons.

The idea of a muon microscope [34,35] was discussed as an active area of development using low-energy muons. The higher the flux of low-energy muons, the greater the ability to tune the beam properties for beam focusing and steering. A high-flux muon source in the US would present the opportunity to design a beamline with these characteristics for measuring textures samples and devices.

Concluding his talk, it was agreed by most participants that LE-µSR represents the most revolutionary advance in µSR recently, as well as the most rapidly growing part of the technique. Its demand will only increase, and it is vitally important that a next-generation µSR facility be able to support world-leading LE-µSR experiments.

### 1.6 STEPHEN BLUNDELL, OXFORD UNIVERSITY – 'THE MUON-FLUORINE INTERACTION: A MODEL QUANTUM SYSTEM FOR EXPLORING DECOHERENCE'

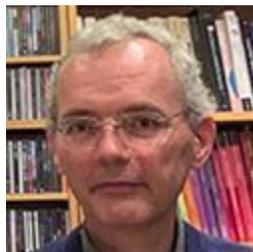

Dr. Blundell is a Professor of Physics in Oxford University, whose research employs µSR to study superconductivity and magnetism in a range of organic and inorganic materials. His recent work developing the technique DFT+µ demonstrates the inarguable benefits of combing advanced computational techniques and muon science.

Dr. Blundell discussed his advances combining computation and µSR in great detail. DFT+µ is a computational technique for µSR that utilizes DFT routines with the addition of a positive muon to the crystal structure. When the muon is added, the structure relaxes, and finds a new ground state configuration around the muon. In this approach, many different muon sites are compared, and the lowest energy configuration is the one that equates to the actual stopping site of the muon. By comparing the energy of that state to other states, you can find the likelihood of having multiple muon stopping sites and/or the energy cost of a muon hopping between sites [36]. As µSR is a local probe, knowledge of the muon stopping site is necessary to extract many physical parameters of the system, meaning that determining the muon stopping site in a material should be seen as a standard component of the experiment planning. Since it can be done offline, it is possible to do a DFT+µ calculation prior to the experiment and use the result to help determine the best way to execute experimental goals. It is most useful when performing data analysis, as it can translate the muon precession frequency into a magnetic moment size and a hopping rate into an energy scale, for example. The DFT+µ technique is gaining more widespread use, but is not yet a standard tool of the field. The timeline for developing a next-generation muon source pairs well with the ongoing efforts in software tools and has the opportunity to incorporate them as standard tools on new beamlines. This is particularly true for a next-generation muon source that incorporates an existing user program and computational scientists.

Dr. Blundell also touched on the unique challenges of a high-flux pulsed muon source when it came to data collection. As with any technique, the key is to be able to collect data as efficiently as possible, and to avoid throwing away any good events. This is especially true for µSR, which is statistically-based, and where subtle information about the material being studied requires large numbers (up to $10^9$) events [37,38]. For a pulsed source with high muon flux, this means having a very large number of detectors and the timing resolution to process millions of events without electronic pileup. It was also suggested that new beam focusing/steering techniques could be used with high data rates in order to enable new experiments. For example, with many detectors it would be possible to raster the beam across the sample and use particle tracking to get information as a function of spatial resolution across a sample. The development of a new US-based muon source would present the opportunity to further develop these enabling technology to drive new scientific discovery.

## 1.7 SARAH DUNSIGER, TRIUMF & SIMON FRASER UNIVERSITY – 'FROM CORRELATIONS TO FUNCTIONALITY USING DEPTH RESOLVED SPIN RESONANCE TECHNIQUES'

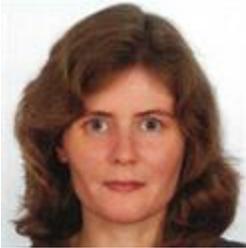

Dr. Dunsiger is a research scientist in TRIUMF's Center for Molecular and Materials Science and Adjunct Professor at Simon Fraser University. Her work using low energy µSR and β-NMR demonstrates the unique power of depth-resolved resonance probes to illuminate the effects of surfaces in quantum materials and interfaces in functional magnetic heterostructures.

She started by establishing the long history of µSR techniques in studying magnetically frustrated systems. In particular, she focused on the role of µSR in revealing the rich magnetic phase diagram of cubic chiral magnets MnSi. This material hosts single Bloch-type skyrmions resulting from competing exchange and Dzyaloshinskii-Moriya interactions. Small angle neutron scattering (SANS) showed that skyrmion formation can be controlled by an applied magnetic field. This has been akin to 0 and 1 bits that makes skyrmion hosting materials possible for applications in data storage and structure. Sarah also presented the landscape of complementarity between various experimental techniques with respect to fluctuation rate: AC susceptibility, NMR, µSR and neutrons (Q resolved). She then discussed how µSR was used to further study the MnSi system in MBE grown thin film. In thin film studies of MnSi samples, the skyrmion phase is stabilized by thermal fluctuations, even if it is energetically unfavorable according to Mean field theory. Enhancement of the low energy excitations in the skyrmion state is consistent with the observed enhancement of the $1/T_{1\,in}$ µSR. For thin films, spin anisotropy becomes more critical and rotating muon polarization makes this investigation possible.

She also identified some open questions where µSR may help. In $Cu_2OSeO_3$ system, Resonant X-ray Scattering (REXS) has identified skyrmion phases but would most likely need µSR to resolve the details. In the canonical spin glass system, AuFe, where depth dependent spin dynamics is of interest, initial µSR measurements were able to show increasing spin fluctuation rate as confinement increased by making the films thinner.

She gave a very interesting example of using µSR to study thermodynamic phase transition in frustrated magnetic metamaterial. These are artificial spin systems that are lab synthesized using electron beam lithography to form kagome lattice of permalloy materials $Ni_{80}Fe_{20}$ with Tc>400K. The system mimics Ising-like spins that shows 2 magnetic transitions in to Ice I and II structures. The µSR showed characteristic muon spin depolarization in kagome ice metamaterials.

In summary, Sarah contends that boundaries and confined geometries often lead to novel behavior making them well suited to being investigated using depth resolved low energy muons and ß-NMR on complimentary time scales. There is great potential to explore systems such as (1) skyrmion lattices (2) geometrically frustrated systems (3) artificial spin systems. In thinking of a future muon spectroscopy facility, she indicated that reasonable timing resolution is of enormous value to understand these systems. Also, because novel materials are often small, a beam spot <5mm across will facilitate experiments in MBE and PLD grown films. In addition. a spin rotator is vital to investigate anisotropy, a major phenomenon in magnetic heterostructures.

Sarah ended her presentation by presenting the many opportunities for collaboration brought about by close complementarity of spin resonance techniques ß-NMR (done at TRIUMF) and bulk µSR at the future µSR facility at the SNS. Furthermore, many opportunities exist to collaborate on thin film synthesis with ORNL's Center for Nanophase Materials Sciences (CNMS).

## 1.8  TRAVIS WILLIAMS, OAK RIDGE NATIONAL LABORATORY – 'SEEMS: A SINGLE EVENT EFFECT AND MUON SPECTROSCOPY FACILITY AT THE SPALLATION NEUTRON SOURCE'

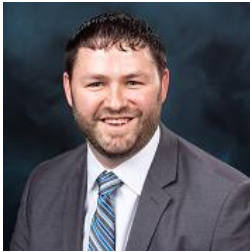

Dr. Williams is an Instrument Scientist within the Neutron Sciences Directorate at ORNL. His work focuses on *f*-electron systems using both neutrons and µSR. He has been involved in developing the science case for the Second Target Station as well as the conceptual design for a muon source at the SNS.

This presentation described the efforts to develop a conceptual design for a next-generation muon facility in the United States. The ongoing upgrades to the Spallation Neutron Source (SNS) at ORNL stimulated discussion about whether the SNS could also support a µSR facility. In 2016, a group of µSR experts from around the world was invited to participate in a small workshop at ORNL to discuss the scientific and technical requirements for a µSR facility. There, a novel approach was found that made use of ORNL-developed laser stripping technology, which could divert a small fraction of the SNS beam to support a future µSR target station. The benefits of this approach was that it could divert the full intensity of the SNS beam at a very low duty cycle, ideal for generating muon beams for µSR, but producing a negligible effect on the beam used to produce neutrons. This would take 0.3% of the SNS beam and would produce up to several orders of magnitude more muon flux than at any comparable facility.

The work done at ORNL on this concept since that time has been to explore the space requirements, the necessary technological developments and produce a baseline cost estimate. A conceptual building layout was presented that suggested that the facility could be constructed for a cost ~$300M (USD), housing an initial suite of 4 muon beamlines with room to support additional beamlines or other missions, including a Single Event Effects (SEE) testing for the avionics industry or particle physics research. The preliminary design presented included beamlines for SEE and Muon Spectroscopy, leading to its acronym, SEEMS. With the feedback generated from this workshop, further refinements of the SEEMS design could be made, including the inclusion of more low-E µSR capabilities, high throughput detectors and micro-focused beams. The µSR community at the workshop expressed a strong desire to continue to be updated on the project and to provide input on the design parameters.

## 1.9  CLARINA DELA CRUZ, OAK RIDGE NATIONAL LABORATORY – 'FUTURE PLANS'

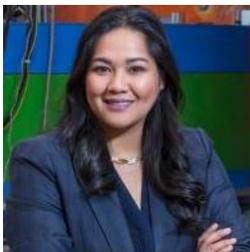

Dr. dela Cruz is the head of the Quantum Materials Initiative within the Neutron Sciences Directorate at ORNL. In this role, she is responsible for evaluating internal research proposals and coordinating external funding opportunities by ORNL staff, including work on a future ORNL muon source.

She discussed the efforts of local scientist and engineers, who have worked together for the past several years to socialize the idea of building a dual-purpose facility housed at the current Spallation Neutron Source (SNS) campus that will deliver capabilities for muon spectroscopy and SEE testing. The main future plans of the team is to collaborate with international muon spectroscopy facilities (TRIUMF, ISIS Neutron and Muons Source, PSI-µSR, JPARC-MLF) to strengthen the scientific and technological community that will be critical to pave the way for support and funding to build the first and state-of-the-art Muons facility in the USA. The team also plans to continue their engagement with prospective industry partners and other stakeholders that will support the proposal for the construction of the most powerful and world leading Single Even Effects (SEE) testing facility at ORNL.

Following the success of the "Inaugural Meeting on the Opportunities of the SEEMS Facility" held on June 11-2, 2019, the team is planning to reconvene for another meeting in 2022. The engagement born out of that workshop has resulted to support letters addressed to ORNL Director's Office, from GE-Aviation, BOEING, Honeywell and Vanderbilt University, expressing their ardent support and need for the SEEMS facility for their business.

The team will aggressively continue engagement with prospective users of the SEEMS facility by delivering talks about the proposed facility in relevant workshops, university seminars and coordinate with Science and Engineering professional societies. The team will work with the SEEMS advisory board to get feedback on the activities being pursued. The team will continue engagement with our targeted funding agencies (ORNL/BES-DOE and NSF). At the same time, the unique characteristics and some first of its kind capabilities delivered by the SEEMS facility will enable the team to pursue new possible stakeholders in industries for computing, microelectronics etc.

## 2. SUMMARY OF BREAKOUT SESSIONS

### 2.1 QUANTUM MATERIALS

Quantum materials are the largest research area for µSR experiments, and are expected to play a large part of the research program for a next-generation muon source. The discussion in this breakout session highlighted the strong complementarity of muon and neutron measurements for quantum materials, and the advantages of a co-located, high intensity muon source with neutron scattering facilities.

The synergy between muons and neutrons in quantum materials was noted in several aspects, including the measurement timescale and elements that could be measured, but much of interest came from the use of muons as a local probe of materials. Neutrons are extremely sensitive to long-range, periodic structures, but muons can provide complementary information to local variations in material properties. Measurements of flux lattices in type-II superconductors were noted as one such area, where muons can effectively measure penetration depths and superfluid densities, even in highly disordered superconductors or with cases of strong pinning effects. The discussion also pointed to magnetically frustrated materials where muons can probe local order and fluctuations, where the local magnetic properties can vary greatly from the bulk magnetism. In both cases, more integration with high-performance computing and DFT modeling would be helpful in determining the muon stopping sites to quantify the results. Participants in the discussion noted that there are several computational efforts underway to develop tools for muon stopping sites and that these tools could be integrated into a next-generation muon source.

The local probe nature of muons was also highlighted in several examples where neutron data could not distinguish between first- and second-order transitions, but muon measurements of the magnetic volume fraction showed them to be clearly first-order transitions. The question was asked whether there are other cases where neutron data has been used to classify a phase transition as second-order may have been due to a lack of local probe measurements. The answer with which most participants agreed was that without complementary measurements it is impossible to even know how frequently this may occur and highlights the need for not relying on a single measurement technique. A co-located muon and neutron source where

### Muons for quantum materials

Shan Wu, et. al. (2021) In preparation.

S. Hameed* et al, In preparation (2021)

- *Muon facility characteristics*
  - Access to a broad range of sample environments (*T*, **B**, *P*, strain)
  - Intense focused beams to probe very small samples
  - A range of energies to tailor deposition depth
  - Spin rotated measurements to probe magnetic anisotropy in single crystals
  - Fast automated turnaround between samples to probe doping sequence
  - Intense ultra short pulses (<50 ns) for phenomena with fast muon relaxation and pump-probe

users have access to both techniques with relative ease would not only prevent such erroneous conclusions, it would illustrate new phenomena that may not be observable by one measurement on its own. It was also realized that order parameter measurements require collecting a high-density of datapoints near the critical point and are excellent candidates for high-throughput characterization. A high-flux muon source would be ideally suited to these measurements.

Muons were also noted as having distinct advantages over neutrons for several key classes of experiments in quantum materials. As a local probe, muons do not require large single crystals and can perform measurements on powders as easily as in crystals. There are many materials for which single crystals do not grow easily or at all, which precludes extracting certain information with neutrons. Incommensurate structures are also more amenable in many ways to local probes, as even small deviations from commensurate structures can be measured. Finally, measurements under pressure are highly sought as pressure is a clean tuning parameter, but measurements in pressure cells require the use of small samples and introduce large backgrounds when measured with neutrons. Using muons, sample size limitations can be overcome, and a high-energy muon source could produce decay muons in large enough quantities to provide tunable muon beams to penetrate pressure cells and only stop in the sample, avoiding additional background signals.

The session was best summarized by a participant who noted that some of the most important observations in quantum materials come with the introduction of chemical doping, which naturally introduce disorder. In order to understand the material, you need to disentangle the effects of randomness from the chemical variation, which requires both local and bulk probes, sensitive to volume fractions and site-averaged information. In this way, the combination of muons and neutrons is an extremely powerful tool for understanding quantum materials, and next-generation measurements are needed in both areas to continue to push scientific boundaries.

## 2.2 CHEMISTRY & BATTERY MATERIALS

Chemistry utilizes µSR measurements in three major areas: hydrogen reaction, muoniated radicals, and spin-labeling. It is used much like electron paramagnetic resonance (EPR), but has distinct advantages. It relies on positive muons capturing an electron to form muonium (Mu). While the muon is much lighter than the proton, the reduced mass of Mu is within 0.5% of hydrogen, making it an excellent substitute in materials. This allows for the study of hydrogen chemistry under conditions that are not available to other techniques. For example, µSR can be used to measure the chemical environment of hydrogen in solid, liquid or gaseous materials. In all of these phases, µSR can be used to determine rate constants as a function of chemical concentration, giving it a very wide range of application compared to other techniques. The main limitation to µSR measurements of hydrogen chemistry is the muon flux. A new muon source with increased flux would have the opportunity to improve the quality of these measurements by increasing the sensitivity, throughput and types of materials available to be measured.

Muon chemistry is also used to study of reactive intermediates, by forming muonium on an unsaturated bond, obtaining a muoniated radical. This is a very clean way to study radicals, as it is an inert process that does not form damaging chemicals or bring about radiolysis. Since the concentration of muonium in the material is very low, it does not induce bimolecular reactions or any secondary/tertiary reaction products. It can also be studied in conditions where radicals are highly mobile, such as at high temperatures. Muoniated radicals are studied by measuring avoided level crossing (ALC) resonances, which require enough closely spaced field measurements to define a lineshape, while having a large enough field range to find resonances. While the field range is dictated by the nature of the measurement, the flux determines the duration of the measurement and the definition of the lineshape. Thus, the higher the flux, the more systems and measurements that can be performed.

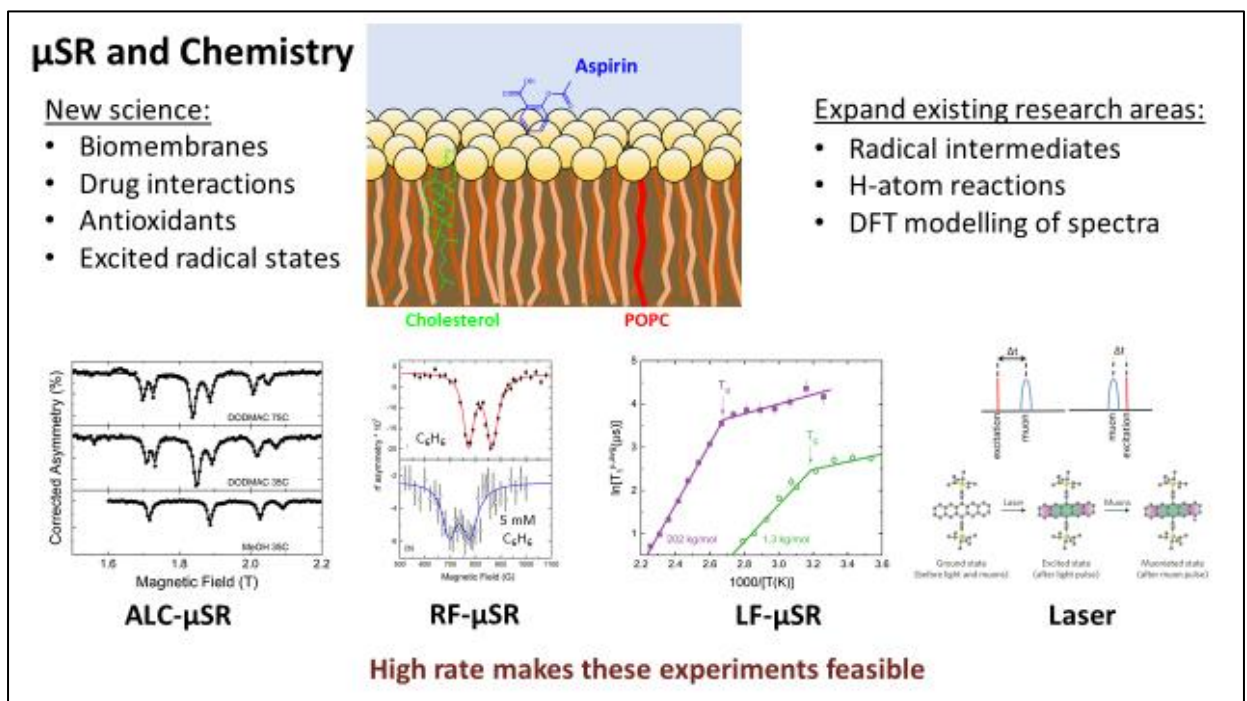

The use of a pulsed source also has advantages for muon chemistry, as it is very amenable to the use of radio frequency (RF)-μSR techniques. Here, the RF pulse can be synchronized to the muon pulse to measure hyperfine coupling constants. Simultaneous measurements can be performed *in situ* in between pulse bunches to provide complementary information. Having co-located laboratory and measurement capabilities will further extend the scale of scientific problems that can be addressed with μSR for hydrogen chemistry.

The ability to perform *in situ* measurements during a μSR measurement are also extremely valuable for measuring battery materials and other forms of ionic diffusion. There was discussion of ongoing work using Python libraries to provide information to aid μSR measurements of chemical systems and battery materials. They aim to identify muon stopping sites, calculate hyperfine couplings and estimate quantum effects on the muon. Continued development of these and other software tools would provide more direction for identifying new materials and phenomena.

In a next generation μSR facility, chemistry and battery materials research would benefit from low-energy muon beams and smaller spot sizes. There is a need to measure surface effects and depth-resolved studies, and often these samples are very small, so focused muon beams would be highly useful. Finally, participants noted that with a high flux source, measurements with an extremely fine resolution pulse (5-10ns) would be achievable for precision measurements.

## 2.3 LOW-ENERGY MUONS

The discussions for the breakout session focusing on low energy muons (LE-μSR) started by identifying the science drivers for this capability. The main science drivers for low energy muons stems out of research in quantum materials (QM). Another driving force is applications in engineering devices where the physics fundamentally changes at the interface. In particular, the use of LE-μSR in characterizing QM for applications, incorporated into device geometries where it is necessary to study the interface properties of the QM. Here, LE-µSR is complimentary with neutrons in identifying interface structures of QM for specialized functionalities. There is a distinct advantage to use LE-µSR with very thin films to see novel states of matter at interfaces of heterostructures composed of magnets, superconductors, topological semimetals, among others, and the possibility of seeing highly exotic states like monopole superconductors. It will also be possible to observe the effects of confinement, proximity broken translation symmetry at interfaces. The capability of LE-µSR to characterize buried interfaces will be game changing! An important characteristic relevant here is the controllability of the LE-µSR, its penetrability, giving it unique access to buried interfaces.

The group expressed that battery materials research need far more LE-µSR compared to QM (~5 muon lifetimes vs QM which only needs ~2 muon lifetimes), which would be made accessible by the pulsed timing structure and increased flux. There are several "green" thin film technologies related to climate change efforts envisioned to include research in thin films for batteries, biofilters and solar cells.

Another science driver identified for LE-µSR is in the research of membranes and diffusion in biological systems. This includes studies of diffusion across biomembranes and development of functional biomembrane devices. There is a possibility for LE-µSR to help where it has not been possible to perform resonance techniques used by biologists and chemists at the LE-µSR instrument at the PSI muon source. For this purpose of performing RF resonance measurements, one needs to measure at longer times. This is the benefit of a pulsed muon source, which has low background in between pulses and can take advantage of orders of magnitude higher muon flux than is currently available. Additionally, RF measurements require a short pulse structure so that the RF power is applied for a small duty cycle to avoid coil and sample heating. Applications of LE-µSR for soft chemistry were also identified as a science driver. In this case, ultra-high vacuum is necessary. Otherwise, the muons will not reach the sample.

The breakout discussion then proceeded to talk about aspects of the instrumentation that will be critical to execute on the science drivers mentioned above. The group agreed that particle physics techniques are standard across existing muon facilities, and there is much to gain from continued synergies between the requirements and capabilities of the various science communities. In this way, LE-µSR truly becomes a new "eye" to look at existing and new materials to see new physics. In studying quantum materials where

## Science drivers for low E muons

- Interface quantum materials, Topological materials
  - Effects of confinement,
  - Proximity broken translation symmetry
  - Surface states
- Advantage with very thin films: heterostructures where magnets, SC's, see novel states of matter at interfaces
- Buried interfaces will be game changing
- Time resolved pump probe or stroboscopic measurements
- Thin films
  - "Green" thin films for climate change efforts
  - Thin films for batteries
  - Biofilters
  - Solar cells
  - Coatings and additive manufacturing

X. Kou et al., Solid State Comm. 215 (2015) 34

magnetism is typically anisotropic, it was identified that a spin rotator will be necessary on the muon instrument. There is interest in doing time resolved pump probe or stroboscopic measurements primarily to investigate QM surface response to pulsed muons. The challenge being it is hard to produce RF pulsed muons. The group identified the need to use Artificial intelligence (AI), use of big data science and large-scale simulations to analyze new muon experimental data. This will require capabilities to model multilayer systems and implement DFT modelling. This will make multimodal experiments possible (muons, neutrons, xrays, NMR etc), help optimize experiments and even experiment autonomous steering in the future. It is also worth noting that AI will allow for better modelling for water in biosystems that may enable using LE-µSR for complex biological samples.

The idea of the muon microscope was also discussed. This entails re-accelerating ultra-LE-µSR to make a beam (~200keV) to focus on a tiny spot. This will only be possible with the projected high intensity of the proposed ORNL muon source. The discussion proceeded to identify research and development needs, where there are technology gaps in the generation of LE-µSR by laser ionization of thermal muonium. One loses flux because thermal muonium in vacuum uses structured aerogel samples and need a strong magnet to keep polarization of muons. There is a need to develop technology to keep the flux of the polarized beam high. There is also a need for high efficiency/power lasers for ionizing muonium, which remains a huge challenge today. There is also the problem of in generating Lyman-α (10 eV) for light amplification. Here, specialized materials with very high purity are needed.

The discussion closed with ideas for what kind of instruments the group wants to see. The very high demand of LE-µSR suggests that 2 out of 4 muon beamlines should be dedicated to LE-µSR. It was identified that instruments should have detectors and electronics that can handle the very high flux of a next-generation muon source. The group insists that sample environments should be considered with the instrument design and not an afterthought.

## 2.4 BROADER IMPACTS

There is a huge need for accelerated neutron testing facility to study Single Event Effects (SEE). The interest is wide ranging, from commercial aircraft industry, military aircraft, drones, automotive systems/autopilot and critical ground-based computing systems. SEE results to unintended behavior, corrupted data, microprocessor/FPGA/ASIC halts and interrupts, unplanned events including equipment failure. High energy neutrons can interact with IC's and cause SEE upsets. SEE is a disturbance of an active electronic device (transistor/gate) caused by energy deposited from the interaction with a single energetic particle. An event occurs when an ionization charge from the energy deposition exceeds the device critical charge.

The risk of SEEs in avionic electronic systems is increasing because: (1) Technology is trending toward smaller features, higher densities and lower voltages resulting in greater susceptibility to atmospheric neutrons (2) number of memory bits and registers are greatly increasing (3) number of flights at higher altitudes is increasing due to better efficiency (4) number of polar flights increasing. There is only one testing facility in the US, the 50 yr old accelerator at LANL that offers high energy neutron testing that matches the atmospheric characteristics. There is a need for a year-round access to high energy neutron beam, thermal energy test capability and systems level test capability which needs large area neutron beams. Additional capacity is also needed to meet current and future needs for avionics and other safety/reliability critical industries. The proposed muon source at ORNL could easily be modified to include neutron beams that are suitable to this purpose.

> # Broader Impacts of muon science
>
> - Negative muons and muonic x-rays at ISIS
>
> - Single Event Effects
>   - Interest from commercial airline industry to understand effects of atmospheric radiation
>     - High energy neutrons → unintended behaviour in electronics
>     - Higher need for testing than ever (smaller IC, polar routes, regulation)
>     - New industries equally interested in muon radiation
>   - Protons directed to muon target will also create neutrons of primary interest to industry over volume large enough for full-device testing; this is game-changing
>
> - Particle physics, muonium-antimuonium
>   - Muons can help solve fundamental problems in particle physics
>     - Flavor violating processes, g-2, muon EDM, muonic hydrogen (proton radius)
>   - Can be performed using muons produced at the proposed SEEMS facility
>
> - Muon Spectroscopy Computational Project
>   - Muon sites, multiple interfaces and more

Muons are not only used to measure properties; they can be used to study the most fundamental problems in particle physics. One of these is the "flavor problem" which looks at the patterns of masses of particles. The flavor problem also deals with unexplained oscillations of particles between different flavors (muonium-antimuoniom conversion). This directly feeds into the bigger question of whether there is physics beyond the Standard Model.

Possible experimental studies with muons include lepton-flavor violating processes, lepton flavor conserving processes and lepton number and lepton flavor violating processes. New physics is also sought from new results of the proton's radius from muonic hydrogen. There is rich science to be explored with Muonium. Muonium exhibits flavor-changing neutral current decays that could provide SM-background free probe of new Physics. It is also a heavy-light state that can exhibit flavor oscillations. These oscillations can probe new Physics without the complications of QCD. New data is needed as the last data on muonium oscillation is still back in 1999. With the capabilities a future facility, ORNL could be a world-leader in fundamental physics studies with muons.

There is an ongoing project aimed at producing and maintaining software tools and methods designed to conjugate muon science with computation and make it easier to interpret muon experiments with full power simulations. The main tools offered are pymuon-suite, a Python library designed to automate many computational tasks such as finding the muon stopping site, and MuDirac, a C++ software to compute the Xray spectra of muonic atoms for elemental analysis experiments. A next-generation muon source offers opportunities to grow the suite of computation and analysis tools for muon spectroscopy.

# CONCLUSIONS

The workshop brought µSR experts and interested members of the scientific community together from around the world to discuss the future of the µSR technique and the scientific drivers for a new muon source in the United States. It was successful in those goals, with world-leading µSR experts discussing the role of muons in quantum materials, radical chemistry, battery materials and elemental analysis. It was further discussed how the field is advancing, from the development of low-energy muon beams, muon microscopy, magnet and pressure cell technology, to developments in software for high-performance computation. Several recurring themes were acknowledged: **(1) the advantage of higher muon fluxes to enable new science, (2) increasing demand for low-energy muon beams, (3) the need for more software tools for muon site determination and analysis, and (4) the role of multi-probe studies combining µSR with neutrons and other spectroscopic techniques**. The following action items need to be pursued to realize a roadmap towards a world-leading µSR facility in the United States.

*The primary need for enabling new science with µSR is access to muon beams of higher flux.* Such a facility would impact all areas of muon science. For quantum materials, higher flux means not only faster throughput, but the ability to measure a longer time window. This is important for the detection of weaker magnetism, which allows for measurements of phenomena such as time-reversal symmetry breaking and other weak magnetic signals. In chemistry, the primary limitation to measuring a compound is the muon flux, so a higher-flux source would open new families of compounds for study with µSR. When measuring avoided level crossing resonances, higher muon fluxes give a greater sensitivity to excitation levels, improving the data quality. Higher muon flux is also useful for more general µSR experiments, for example when decay muons are used to measure encapsulated samples for biochemistry or inside pressure cells for parametric studies. Even broader applications, such as using muon beams for fundamental physics experiments, would benefit from the increased data rates of a higher flux source. But the participants were in agreement that the largest benefit of a high muon flux would be the expansion of low-energy µSR capabilities.

*The µSR community strongly agrees that any opportunity to expand the low-energy µSR capabilities be aggressively pursued.* The number of experiments utilizing low-energy µSR beams has continued to expand, and the push to study surface effects and devices will increase the demand. Low-energy µSR beams have the ability to be stopped in very short distances within a sample (on the order of nanometers), but at the cost of several orders of magnitude in muon flux. This has limited the types of experiments that can be done and requires longer experiments to perform the measurements. Doing so would enable a greater portfolio of depth-resolved experiments, meaning more opportunities for measuring surface effects such as topological materials, novel states that form in interfaces, penetration depths in type-I superconductors, as well as layered heterostructures and substrates. With a higher flux of low-energy muons, the possibility of a muon microscope emerged as a new capability that does not currently exist, allowing for these measurements over different spatial regions, for example in different parts of a nanostructured device.

*To aid the planning, execution and analysis of µSR experiments, more software tools are needed.* This was highlighted by several recent developments in both lineshape analysis and muon site determination using DFT techniques. Several participants also shared their work on current software developments in these areas, demonstrating the commitment by the wider community to help quantify µSR data and develop more firm conclusions from the data. More work is clearly needed, and a µSR software suite running on high-performance computing clusters at a next-generation facility would expand the user community to more casual users or researchers with other techniques, thus generating more impactful science. Particularly making muon site determination more routine, as well as tools for calculating muon diffusion or modeling disorder, would be a great improvement for analyzing µSR experiments.

*A next-generation muon source in the US would miss scientific opportunities by not being closely associated with neutron scattering, computing, advanced materials characterization, and other spectroscopic techniques.* A large portion of the discussion at the workshop centered around the complementary nature of µSR to other techniques, especially neutron scattering. In many areas, most notably in quantum materials, it was repeatedly highlighted how the two techniques work together to provide more information on a given system than either alone. Several examples were shown where exclusively using neutrons provided ambiguous results that were clarified with a later µSR experiment. It was noted that there is a danger in relying on any one technique to address a scientific problem, and that researchers always benefit from partnering with or having access to other types of measurements. The fact that 3 out of 4 µSR sources worldwide are co-located with neutron scattering facilities demonstrates this approach, and all three of those facilities see many users that utilize both techniques. Co-locating these facilities and fostering collaborations would be a great benefit to scientific discovery.

*There is a strong desire for the community to remain engaged and to realize new scientific opportunities that could be enabled with a new, US-based muon source.* At the end of the workshop, a concept was presented for a US-based µSR facility located at ORNL's Spallation Neutron Source. While still very early in the process, the participants were encouraged by the work being done to realize this facility. There was a great deal of discussion of its potential capabilities, particularly what the achievable flux of this source could be. It was felt that with the opportunity to gain orders of magnitude more flux, effort needed to be invested to determine these properties more concretely and refine the possible beamline concepts. All the participants felt that this would be an excellent opportunity to develop a US-based µSR facility and would be best positioned to develop new low-energy µSR beams that can push the capabilities that are currently available.

# APPENDIX A. AGENDA

| | | |
|---|---|---|
| **Scientific Opportunities for a US-based Muon Spectroscopy Facility** ||| 
| **Day 1 - February 1, 2021** |||
| 8:00 am - 8:15 am | Welcome & Opening Remarks | Despina Louca<br>University of Virginia |
| | Dr. Louca is the Maxine S. and Jesse W. Beams Professor of Physics at the Univeristy of Virginia. She is a neutron scatterer by training and is currently serving as the president of the Neutron Scattering Society of America (NSSA). Dr. Louca's research is in materials physics, focusing on understanding the interactions of spin, charge and lattice degrees of freedom, and of their role in phase transitions observed in strongly correlated electron solids that exhibit properties such as magnetoresistance, superconductivity and multiferroicity. She is a fellow of the American Physical Society. Dr. Louca sees immense value in bringing muon spectroscopy in the US and is spearheading this effort to get not just the muon community involved but the neutron community as well in making this happen. ||
| 8:15 am - 9:05 am | Quantum Materials with Muons and Neutrons | Tomo Uemura<br>Columbia University |
| | Dr. Uemura is a Professor at Columbia University and the inaugural recipient of the Toshimitsu Yamazaki prize for outstanding, sustained work in µSR with long term impact on science. His foundational work in the areas of spin glasses, quantum magnetism and unconventional superconductivity give him a unique perspective on the power of muon techniques and their impact on condensed matter research. ||
| 9:05 am - 9:55 am | Battery Materials with Muons | Jun Sugiyama<br>CROSS-Tokai |
| | Dr. Sugiyama has a long history of using muon beams for industry-oriented research into battery materials. As a former researcher with Toyota Central R&D Laboratories, he has been a premier expert in measuring lithium and sodium ion diffusion rates in batteries. He brings tremendous insight into the industry motivations for a next-generation muon source. ||
| 9:55 am - 10:10 am | *Break* ||
| 10:10 am - 11:00 am | Muon Spin Relaxation in Semiconductors and Oxides | James Lord<br>ISIS Neutron and Muon Source |
| | Dr. Lord is an instrument scientist on the HiFi muon Instrument at the ISIS Neutron and Muon Source. His research focuses on condensed matter systems, specifically semiconductors and oxides. Dr. Lord is also an expert on muon instrumentation and research using high magnetic field measurements using muon spectroscopy. ||
| 11:00 am - 11:50 am | Ionic Diffusion | Peter Baker<br>ISIS-RAL |
| | Dr. Baker is an instrument scientist at the ISIS Neutron and Muon source, which hosts pulsed muon beams for performing a wide range of scientific experiments. Dr. Baker's research combines probing ionic diffusion with magnetism and superconductivity, which highlights the versatility of muon spectroscopy as a technique. ||
| 11:50 am - 12:30 pm | *Lunch* ||

| 12:30 pm - 1:30 pm | Breakout: Quantum Materials | Collin Broholm<br>Johns Hopkins University |
|---|---|---|
| | \multicolumn{2}{l\|}{Quantum Materials are the largest research area for µSR around the world, and the development of a next generation muon source will open up new research directions. This session will discuss possible avenues for new research in frustrated materials, topological states, low-dimensional systems and beyond.} |
| 1:30 pm - 2:30 pm | Breakout: Chemistry & Battery Materials | Iain McKenzie<br>TRIUMF |
| | \multicolumn{2}{l\|}{µSR has been a unique probe for studying ionic diffusion, catalysis and battery materials. For these applications, pulsed µSR sources are ideal due to their higher flux. This session will discuss how a next-generation source with orders of magnitude increases in flux can facilitate new research directions.} |
| 2:30 pm | \multicolumn{2}{c\|}{*Adjourn*} |

| | | |
|---|---|---|
| | **Day 2 - February 2, 2021** | |
| | | |
| 8:00 am - 8:50 am | Low-Energy Muons | Thomas Prokscha<br>Paul Scherrer Institute |
| | \multicolumn{2}{l\|}{Dr. Prokscha is the Head of the Low-Energy Muon group at PSI, the world's best source for low energy muon beams. He brings an unparallelled vision for the development and future need for studies using low-energy muon beams, including to thin films, devices and surface studies.} |
| 8:50 am - 9:40 am | Muons and Computation | Stephen Blundell<br>Oxford University |
| | \multicolumn{2}{l\|}{Dr. Blundell is a Professor of Physics in Oxford University, whose research employs µSR to study superconductivity and magnetism in a range of organic and inorganic materials. His recent work devloping the technique DFT+µ demonstrates the inaruguable benfits of combing advanced computational techniques and muon science.} |
| 9:40 am - 10:00 am | \multicolumn{2}{c\|}{*Break*} |
| 10:00 am - 10:50 am | LE-uSR and B-NMR | Sarah Dunsiger<br>TRIUMF/Simon Fraser U. |
| | \multicolumn{2}{l\|}{Dr. Dunsiger is a research scientist in TRIUMF's Center for Molecular and Materials Science and Adjuct Professor at Simon Fraser University. Her work using low energy µSR and beta-NMR demonstrates the unique power of depth-resolved resonance probes to illuminate the effects of surfaces in quantum materials and interfaces in functional magnetic heterostructures..} |
| 10:50 am - 11:50 am | Breakout: Low-Energy Muons | Alannah Hallas     &     Alan Tennant<br>Univ. of British Columbia     Oak Ridge National Lab |
| | \multicolumn{2}{l\|}{Low-energy muon beams are a rapidly-growing area of µSR research, due to their ability to probe surface states and depth-resolved properties. Currently, low-E µSR beams have been limited by low flux, but a new µSR facility has the capability to overcome this shortcoming. This session will discuss the scientific needs for low-E µSR capabilities in a future US facility.} |
| 11:50 am - 12:30 pm | \multicolumn{2}{c\|}{*Lunch*} |
| 12:30 pm - 1:30 pm | Breakout: Broader Impacts | Ady Hillier     &     Tom Lancaster<br>ISIS-RAL     University of Durham |

|  | The benefits of a high-flux domestic muon source extend beyond the applications of μSR. This session will focus on the broader impacts of a next-generation muon source, including integration with SEE testing and fundamental physics.  It will also discuss new developments within the user community and opportunities for growth. |  |
|---|---|---|
| 1:30 pm - 2:00 pm | Summary of Breakouts | Greg MacDougall<br>University of Illinois |
|  | Dr. MacDougall is a Research Professor at the University of Illinois at Urbana-Champaign. He specializes in the growth and characterization of novel quantum materials, and has years of experience utilizing μSR and neutron scattering instrumentation at National Laboratories. In recent years, he has been working with ORNL to help develop plans for the next-generation SEEMS muon facility. |  |
| 2:00 pm - 2:30 pm | SEEMS Presentation | Travis Williams<br>Oak Ridge National Lab |
|  | Dr. Williams is an Instrument Scientist within the Neutron Sciences Directorate at ORNL. His work focuses on f-electron systems using both neutrons and μSR.  He has been involved in developing the science case for the Second Target Station as well as the conceptual design for a muon source at the SNS. |  |
| 2:30 pm - 2:45 pm | Future Plans | Clarina dela Cruz<br>Oak Ridge National Lab |
|  | Dr. dela Cruz is the head of the Quantum Materials Initiative within the Neutron Sciences Directorate at ORNL.  In this role, she is responsible for evaluating internal research proposals and coordinating external funding opportunities by ORNL staff, including work on a future ORNL muon source. |  |
| 2:45 pm | *Adjourn* |  |

# APPENDIX B.  COMPARISON OF EXISTING MUON SOURCES WORLDWIDE

| Facility | Source Type | Flux ($\mu^+$/sec)* | Pulse Width (ns) | Beamlines§ |
|---|---|---|---|---|
| ISIS | Pulsed | $1.5 \times 10^6$ | 70 | 3 surface / 4 decay |
| J-PARC | Pulsed | $1.8 \times 10^7$ | 100 | 2 surface / 2 decay / 2 low-E / 1 high-E |
| PSI | Continuous | $2.0 \times 10^8$ | - | 4 surface / 1 decay |
| TRIUMF | Continuous | $5.0 \times 10^6$ | - | 3 surface / 1 decay |
| Proposed at the SNS | Pulsed | $7.7 \times 10^8$ | 50 | 2-3 surface / 1-2 low-E / 1-2 decay |

| Facility | Temperature Range | Max Field | Max Pressure | Co-located neutrons? |
|---|---|---|---|---|
| ISIS | 0.03 – 1500 K | 5.0 T | 0.7 GPa | Yes |
| J-PARC | 0.05 – 1000 K | 3.5 T | 1.0 GPa | Yes |
| PSI | 0.01 – 1000 K | 9.5 T | 2.8 GPa | Yes |
| TRIUMF | 0.02 – 900 K | 9.0 T | 0.5 GPa | No |
| Proposed at the SNS | - | - | - | Yes |

\* Flux represents the time-averaged flux per instrument, averaged across the facility.  Note that continuous sources operating in time-differential mode (measuring one muon at a time) are thus rate-limited to $7.0 \times 10^4$ $\mu^+$/sec.

§ Of the currently-constructed beamlines, the decay muon beamline and one of the surface muon beamlines at TRIUMF, as well as the decay muon beamline at J-PARC, are not currently operational.